\documentclass[12pt]{article}
\usepackage[dvips]{graphicx}
\begin{document}
\date{\today }
\title{Multiscale fluctuations in nuclear response }
\author{D.~Lacroix, Ph.~Chomaz \\
\medskip (1)G.A.N.I.L., B.P. 5027, 
F-14076 Caen Cedex 5, France.}
\maketitle

\begin{abstract}
The nuclear collective response is investigated in the framework of a
doorway picture in which the spreading width of the collective motion is
described as a coupling to more and more complex configurations. It is shown
that this coupling induces fluctuations of the observed strength. In the
case of a hierarchy of overlapping decay channels, we observe Ericson
fluctuations at different scales. Methods for extracting these scales and
the related lifetimes are discussed. Finally, we show that the coupling of
different states at one level of complexity to some common decay channels at
the next level, may produce interference-like patterns in the nuclear
response. This quantum effect leads to a new type of fluctuations with a
typical width related to the level spacing.
\end{abstract}

\section{Introduction}

\smallskip

Lifetimes and damping mechanisms are general questions in physics. In
quantum mechanics, lifetimes are directly related to the width of the
considered states through the Heisenberg relation. Lifetimes of simple
systems, such as single particle states are often related to quantum
tunnelling. In many-body systems, the damping of excitations appears{\bf \ }
more complex since it can involve many different processes. This is in
particular the case for collective excitations of nuclei for which several
origins for the observed width have been discussed so far\cite
{Ber83,Ber94,Cho95}. On the first hand, a fragmentation of the collective
response into several collective states is expected, leading to the
so-called Landau damping. The collective motion can also be directly coupled
to the continuum of escaping states\cite{Mol68,Kle85}, giving rise 
to the escape width. Finally, the collective strength can decay toward 
the compound nucleus
configurations. This spreading width is often viewed in a doorway
picture as the coupling to more and more complex states. For instance, a
collective excitation built from particle-hole excitations can be coupled to
two-particle two-hole states through the residual two-body interaction
\cite{Dro90}. Again, those states might themselves decay toward
three-particle three-hole states because of two-body collisions. After many
such steps up in complexity, this process eventually ends in the chaos of
compound nucleus states\cite{Cho94,Cho95-2,Lau95,Zel96}( which may, in general, induce a
fine-structure in the response of the nucleus\cite{Win83}). In such a
picture, collective excitations of many-body systems exhibit a large variety
of time scales for the decay mechanism going from the short lifetime of the
collective motion associated to a width of several MeV, to the long lived
compound nucleus states with a typical width of the order of few eV or even
less.

\smallskip

Correlations are a well known tool to measure lifetime of quantal system.
Indeed, when the level spacing is small compared with the typical width of
each individual level the observed spectrum is known to exhibit Ericson
fluctuations\cite{Eri63}. In such a case, excitation spectra presents
fluctuations with a correlation width characteristic of the average width of
the overlapping states. As far as the collective response is concerned,
because of the various scales involved in the decay process, one may wonder
about the characteristics of the expected fluctuation pattern. Since the
compound nucleus states can be considered as the {\bf \ }true{\bf \ }%
eigenstates of the many-body system, one would predict that the observed
fluctuations should be characteristic of the associated lifetime. However,
these fluctuations are over a so small scale that most experiments are
unable to detect them. On the other hand, collective states have a large
width and indeed poor resolution experiments exhibit resonant structure with
several MeV width. Improving upon the resolution one thus expects to uncover
more and more detailed structures directly related to the different levels
of complexity in the damping mechanism. The Landau spreading can be first
studied. Then, according to the commonly admitted picture, one may look at
the coupling to the two-particle two-hole states using a finer coarse
graining, and so on and so forth down to the compound nucleus scale. Is it
possible to observe such a multiscale structure in the fluctuation pattern,
from the collective mode toward different levels of complexity down to the
compound nucleus chaos? In such a context, the response function may present
more and more detailed structure when the resolution is improved, so one may
even ask himself about a possible fractal nature of the response function.
In this case, are the fluctuations self-similar or not? And so which
interpretation can be given to the observed fluctuations as a function of
the resolution in used?

However, the above discussion might well be too simple. Indeed, the
robustness of the various complexity levels against the effects of the
residual interaction, has not been investigated. Even if from the classical
point of view a typical path toward chaos may go through a sequence of
bifurcations, there is no guarantee that, in quantum mechanics, the coupling
of a regular collective motion to an ensemble of chaotic states follows this
path. The simple picture of a hierarchy of more and more complex doorway
states is typical of a perturbative approach, but in such large matrix, one
strong element (compared with the typical level spacing) may be enough to
change the spectrum and the properties of many eigenstates: i.e., at some
point, the perturbative approach may break down. Moreover, many quantum
effects can also be expected such as the possibility of interferences
\cite{Mah69,Sok89,Sok97-2} or the
indirect coupling of doorway states through their decay at higher orders in
perturbative expansions.

In this article we would like to present a critical discussion of the
various questions raised above. The possibility of having multiscale
fluctuations is first illustrated in section II. Since, it might be
difficult to disentangle various scales in a fluctuating spectrum, we
present in section III a brief discussion of the various methods for
extracting these scales. Finally, in section IV, we demonstrate through
analytical derivations and numerical calculations that fluctuations may also
occur due to quantum interferences.

\smallskip

\section{Formalism and results}

\smallskip Let us first briefly recall some known results about damping
mechanism of doorway states and about the fluctuations of cross sections. In
such a way, we will make clear the concepts and notations used in the
article.

\subsection{Ericson Fluctuations in doorway processes}

\smallskip The collective response can be tested by applying an external
field to the nucleus. If we assume that the ground state $\left|
0\right\rangle$ is excited through a time-dependent field $\lambda \hat{D}%
\;\left( e^{-iEt}+\;e^{+iEt}\right) $, where $\hat{D}$ is a hermitian
operator. The linear response theory tells us that the state of the system is%
\footnote{%
in all the paper we used $\hbar =1$}\cite{Rin81} 
\begin{equation}
\left| \Psi \left( t\right) \right\rangle =\left| 0\right\rangle +\lambda
\sum_\mu \left| \mu \right\rangle \left( \frac{\left\langle \mu \right| \hat{%
D}\left| 0\right\rangle }{\left( E-E_\mu \right) +i\Gamma _\mu /2}\;e^{-iEt}-%
\frac{\left\langle \mu \right| \hat{D}\left| 0\right\rangle }{\left( E+E_\mu
\right) -i\Gamma _\mu /2}\;e^{+iEt}\right) \;  \label{eq:1}
\end{equation}
where we have assumed that each eigenstate $\left|\mu \right>$ of the system
has a finite lifetime\footnote{
In the following, the states $|\mu>$ are considered as eigenstates of
an effective hamiltonian. In the first part of this article the coupling
of these 
degrees of freedom to the rest of the system will be simply taken into
account as a  finite lifetime $\Gamma_\mu$. This is valid only for
nonoverlapping decay channels as discussed in refs. \cite{Mah69,Sok89}
and in the last chapter.} $\tau _\mu =1/\Gamma _\mu$. 
Using the doorway notations 
$\hat{D}\left| 0\right\rangle \equiv \left| Coll\right\rangle $ and
performing a Fourier transform of the observation
$\left\langle \Psi \left( t\right) \right| \hat{D}\left|
\Psi \left( t\right) \right\rangle $, 
we get the corresponding spectral response which reads for positive energies 
\begin{equation}
R\left( E\right) =\sum_\mu \left( \frac{\left| \left\langle \mu \right.
\left| Coll\right\rangle \right| ^2}{\left( E-E_\mu \right) +i\Gamma _\mu /2}%
-\frac{\left| \left\langle \mu \right. \left| Coll\right\rangle \right| ^2}{%
\left( E+E_\mu \right) +i\Gamma _\mu /2}\right)  \label{resp}
\end{equation}
Since the first term is dominant we can write $\displaystyle
R(E) \simeq \sum_\mu \left| \left\langle \mu \right. \left|
Coll\right\rangle \right| ^2 / \left( E-E_\mu +i\Gamma _\mu /2\right)$.
The associated strength function is given by: 
\begin{equation}
S\left( E\right) =-\frac 1\pi Im\left( R(E)\right) \simeq \sum_\mu \frac{%
\frac{\Gamma _\mu }{2\pi }\left| \left\langle \mu \right. \left|
Coll\right\rangle \right| ^2}{\left( E-E_\mu \right) ^2+\Gamma _\mu ^2/4}
\end{equation}

When the various states $\left| \mu \right\rangle $ are well separated, i.e.
when the average distance $\Delta E$ between two states is much larger than
the individual width $\Gamma _{\mu },$ the strength $S\left( E\right) $
presents isolated peaks with a typical width $\Gamma _{\mu }.$ When the
states are strongly overlapping, one can still extract information about the
individual widths by looking at the fluctuations known as Ericson
fluctuations\cite{Eri63}. Indeed, the overlap $O_{\mu }=\left| \left\langle
\mu \right. \left| Coll\right\rangle \right| ^{2}$ can be separated into two
parts

\begin{itemize}
\item  an averaged part $\bar{O}\left( E_{\mu }\right) $ which is a smooth
function of $E_{\mu }$. In the case of a resonance, this averaged part often
take the shape of a gaussian or a lorentzian centered on the collective
energy with a width interpreted as the width of the doorway state\cite{Boh70}
;

\item  a fluctuating one $\delta O_{\mu }=O_{\mu }-\bar{O}\left( E_{\mu
}\right) $.
\end{itemize}

Then the response can also be splitted into two components (see appendix 1)
. On the one hand, it contains an averaged strength function $\bar{S}\left(
E\right) $ proportional to $\bar{O}$ and on the other hand remains a
fluctuating term $\delta S\left( E\right) =S\left( E\right) -\bar{S}\left(
E\right)$ associated with $\delta O_\mu .$ These fluctuations still contain
information about the averaged width $\overline{\Gamma }$ as it was
demonstrated by Ericson\cite{Eri63}. To measure fluctuations, we can define
the autocorrelation function of the strength distribution as 
\begin{equation}
C(E)=\frac 1{\delta E}\int_{E_0}^{E_0+\delta E}dE^{\prime }\;\delta S\left(
E^{\prime }\right) \;\delta S\left( E^{\prime }+E\right)
\end{equation}
where $\delta E$ is the averaging interval. If we assume random correlations
between the overlap fluctuation, i.e. $\delta O_\mu \delta O_\nu =\delta
O_\mu ^2\;\delta _{\mu \nu }$, we get the following correlation 
\begin{equation}
C(E)=\frac 2{\pi \Delta E}\frac{\overline{\delta O^2}}{E^2+\overline{\Gamma }%
^2}
\end{equation}
where $\Delta E$ is the averaged level spacing. This relation shows that the
width of the autocorrelation function is directly related to the typical
width of the individual states.

\begin{figure}[h]
\begin{center}
\includegraphics*[height=14cm]{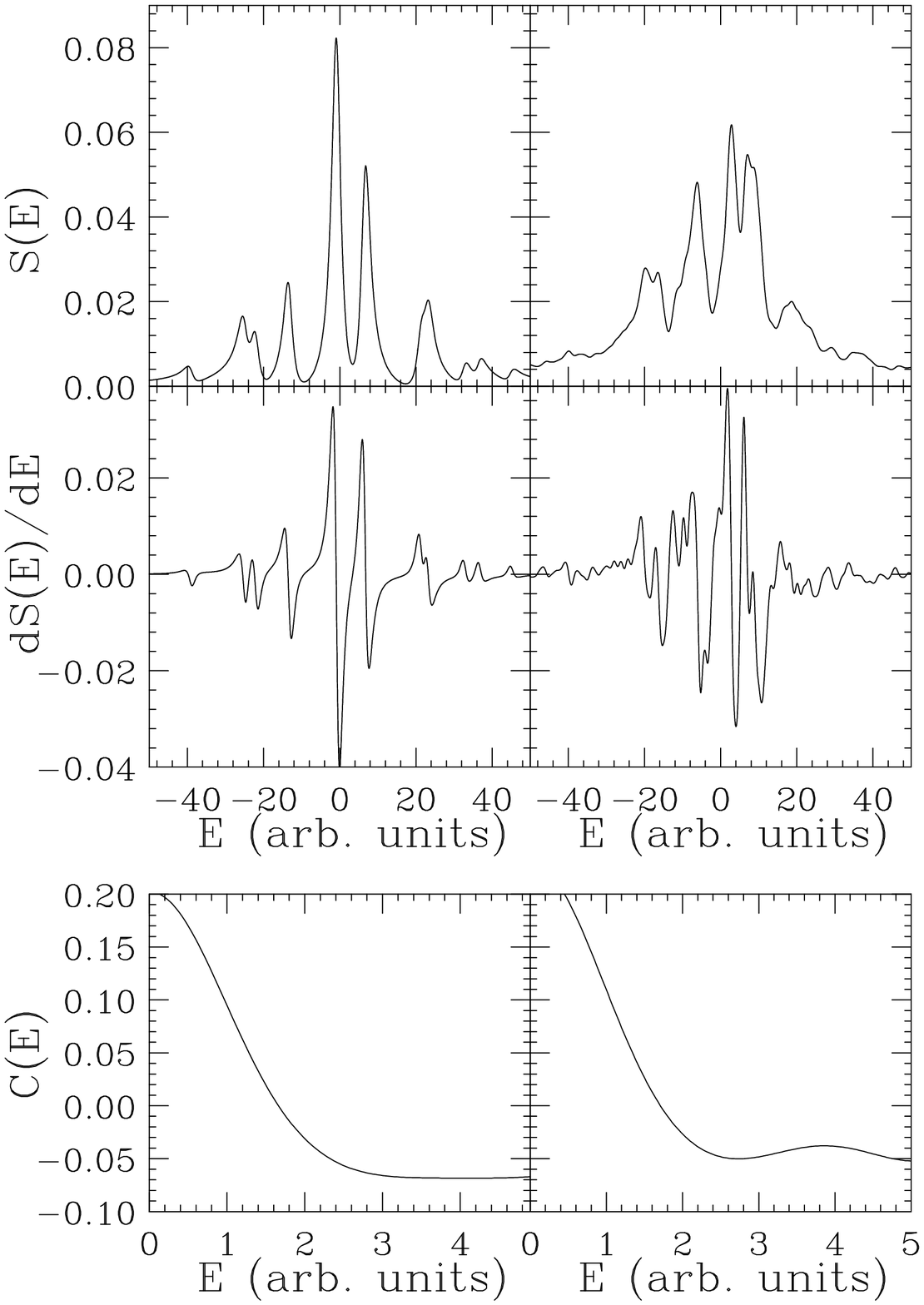}
\end{center}                                     
\caption{Top: Illustration of the strength computed for a collective
excitation coupled to an ensemble of states (decay channel). In left part,
few states with a width $\overline{\Gamma }=3$ smaller than the spacing ($%
\Delta E\simeq 10$) is used. In right part, the opposite case of a dense
spectrum of decay channels ($\Delta E\simeq 0.5$) with an average width $%
\overline{\Gamma }=3$ greater than the spacing exhibit typical Ericson
fluctuations. In both cases, the collective state is assumed to be spread,
giving a variance around $30$ in the considered interval. Middle: derivative
of the strength function. Bottom : auto-correlation performed on the
derivative of the strength function. In both cases, the averaged width of
the decay channel $\overline{\Gamma }$ can be extracted. Note that half of
the width is measured on the autocorrelation of the derivative of the
strength function.}
\label{fig:1}
\end{figure}

In figure (\ref{fig:1}), 
two cases are displayed: in the left part, we illustrate a case with $%
\overline{\Gamma }\ll \Delta E$, which leads to the fragmented response (in
some cases, this is comparable to the Landau spreading), whereas in the
right part, we show a typical Ericson case with $\overline{\Gamma }\gg
\Delta E.$\ On top of this figure, we show the strength function\footnote{%
It should be noticed that, since the energy units can be re-scaled, 
we have chosen arbitrary units to display the various strength functions.}. As discussed in
section 3, instead of performing the autocorrelation technique on the
strength function, it appears more convenient to perform it on the first
derivative of the strength function. Indeed, the autocorrelation on the
strength function is particularly suitable when it is applied to a rather
flat average spectrum. However, in general because of the global variation
of the strength, an average strength has to be removed from the
distribution in order to extract the fluctuations\cite{Kil87}. In the case
of a resonant average response, we have observed that the width extracted
from the autocorrelation procedure is sensitive to the method used to define
this average (fitting procedure, averaging interval). This might introduce
some spurious fluctuations. A way to remove it, is to consider the
derivative of the studied spectrum. % to suppress its global variations.  
The derivative of the strength function emphasizes its fine structure and
avoid the ambiguities on the averaged strength distribution. In the middle
of figure (\ref{fig:1}), we show the derivative of the strength function and
the associated autocorrelation in the bottom. In both cases the width of the
autocorrelation is directly proportional to the width of the individual
states. 
Although the two considered strength function are different, we can 
observe in figure (\ref{fig:1}) that the associated autocorrelation
functions are almost identical. This comes directly from the fact 
that the autocorrelation applied to the derivative is only sensitive
to the smallest scale in the spectra and ignores long-range correlation.
In this figure, since we have imposed the same small fluctuations
scale in both examples, the two autocorrelation happen to be very similar 
showing that this technique could be applied on a fragmented strength or 
a Ericson fluctuation case with the same success.

\subsection{Damping of Collective States}

In order to get a deeper insight into the fluctuation pattern in the
collective response and into the information which can be extracted from
their study, we can now consider the microscopic mechanism responsible for
this damping. The coupling of 
the collective motion to the compound nucleus states leads 
to an internal mixing while the relaxation to the continuum is a
true external decay channel.   
In reference \cite{Sok97}, the effect of the continuum was carefully 
investigated but no particular assumption on the internal degrees of
freedom is made. In the present paper we will mainly focus on the
complexity of the internal mixing. In particular, we will introduce a
hierarchy of degrees of freedom and couplings.

\subsubsection{\protect\smallskip Modelisation of the Doorway Mechanism}

Let us first introduce a doorway state $\left| Coll\right\rangle $
associated with an unperturbed energy $\hat{H}_{0}\left| Coll\right\rangle
=E_{Coll}\left| Coll\right\rangle .$ This state is in general not an
eigenstate of the total hamiltonian but is coupled to more complexe states.
Very often it is possible to introduce a hierarchy of complexity between
those states. In particular, giant resonances are assumed to decay toward
two-particle two-hole states which are themself coupled to three-particle
three-hole states which are damped by more complex states. Such a mechanism
can be described by considering a basis which can be formed into a hierarchy
of states $\left| Coll\right\rangle ,$ $\left\{ \left| i_{1}\right\rangle
\right\} ,$ $\left\{ \left| i_{2}\right\rangle \right\} ,$ $\ldots $ at
unperturbed energies $\hat{H}_{0}\left| i_{n}\right\rangle =E_{i_{n}}\left|
i_{n}\right\rangle .$ These states allows the definition of more and more
complex Hilbert spaces ${\cal E}_{0}\subset {\cal \ }{\cal E}_{1}\subset 
{\cal \ }{\cal E}_{2}$ $\subset $\ldots  where ${\cal E}_{0}$ contains only $%
\left| Coll\right\rangle $ while ${\cal E}_{n}$ includes all the states up
to the $n^{th}$ level, $\left| i_{n}\right\rangle $ . We can now introduce a
hierarchy of residual interactions $\hat{V}_{1}$, $\hat{V}_{2},$ \ldots
where the interaction $\hat{V}_{n}$ couples the states of the space ${\cal E}%
_{n-1}$ to the states $\left\{ \left| i_{n}\right\rangle \right\} .$ Then,
the diagonalization of the Hamiltonian 
\begin{equation}
\hat{H}_{n}=\hat{H}_{0}+\hat{V}_{1}+\ldots +\hat{V}_{n}
\end{equation}
produces eigenstates $\left| \mu _{n}\right\rangle $ 
\begin{equation}
\left| \mu _{n}\right\rangle =c_{\mu _{n},Coll}^{n}\left| Coll\right\rangle
+\sum_{i}c_{\mu _{n},i}^{n}\left| i\right\rangle
\label{wave7} 
\end{equation}
with an energy $E_{\mu _{n}}.$ In the following we will truncate the
hierarchy at various level and discuss the fluctuations properties of the
associated strength.

\subsubsection{The escape width $\Gamma ^{\uparrow }$ as a coherent decay of
the states $\mu $}

At a given level of complexity, the states $\mu $ have an infinite lifetime.
However, in nuclei, eigenstates are coupled to the continuum leading to a
finite lifetime $\Gamma _{\mu }$ for each state. This decay to the continuum
is also linked to the escape width $\Gamma ^{\uparrow }$ of the collective
mode. Indeed, introducing the continuum states $\left| k\right\rangle $ and
a coupling with the states $\mu $ through a residual interaction $\Delta 
\hat{V}$, we can estimate the width $\Gamma _{\mu }$ which is related to the
transition matrix elements $\left| \left\langle k\right| \Delta \hat{V}%
\left| \mu \right\rangle \right| ^{2}$ from the states $\left| \mu
\right\rangle $ to the continuum states $\left| k\right\rangle $. Using the
wave function (\ref{wave7}), we get 
\begin{equation}
\left| \left\langle k\right| \Delta \hat{V}\left| \mu \right\rangle \right|
^{2}=\left| c_{\mu ,Coll}\left\langle k\right| \Delta \hat{V}\left|
Coll\right\rangle +\sum_{i}c_{\mu ,i}\left\langle k\right| \Delta \hat{V}%
\left| i\right\rangle \right| ^{2}
\end{equation}
Since $c_{\mu ,Coll}$ and the various $c_{\mu ,i}$ are not correlated, the
previous expression can be evaluated as an incoherent sum 
$\displaystyle
\left| \left\langle k\right| \Delta \hat{V}\left| \mu \right\rangle \right|
^{2}\simeq \left| c_{\mu ,Coll}\right| ^{2}\left| \left\langle k\right|
\Delta \hat{V}\left| Coll\right\rangle \right| ^{2}+\sum_{i}\left| c_{\mu
,i}\right| ^{2}\left| \left\langle k\right| \Delta \hat{V}\left|
i\right\rangle \right| ^{2}$,
showing that, as far as the decay to the continuum is concerned, taking
advantage of the Fermi golden rule, we can write 
\begin{equation}
\Gamma _{\mu }=\left| c_{\mu ,Coll}\right| ^{2}\Gamma ^{\uparrow
}+\sum_{i}\left| c_{\mu ,i}\right| ^{2}\Gamma _{i}^{\uparrow }
\label{eq:width}
\end{equation}
Here we have introduced the direct coupling of the collective states to the
continuum: the escape width $\Gamma ^{\uparrow }\propto \left| \left\langle
k\right| \Delta \hat{V}\left| Coll\right\rangle \right| ^{2}$. $\Gamma _{\mu
}$ can thus be splitted into two parts: a collective one $\Gamma _{\mu
}^{Coll}=\left| c_{\mu ,Coll}\right| ^{2}
\Gamma ^{\uparrow }=O_{\mu }\Gamma^{\uparrow }$, which is distributed 
according to a lorentzian shape (cf eq. (\ref{usual})), 
and an individual coupling to the continuum $\Gamma _{\mu
}^{\uparrow }=\sum \left| c_{\mu ,i}\right| ^{2}\Gamma _{i}^{\uparrow }$
which is not expected to present any particular structure. Since the $\left|
c_{\mu ,Coll}\right| ^{2}$ are very small, the $\Gamma _{\mu }$ might be
much smaller than $\Gamma ^{\uparrow }$. This is actually the case since
typical values are the eV for $\Gamma _{\mu }$ and the hundreds of keV for $%
\Gamma ^{\uparrow }$. Note that, if we take $\left| c_{\mu ,Coll}\right|
^{2} $ $\propto 1/N$ where $N$ is the number of states coupled to $\left|
Coll\right\rangle $: $N\simeq \Gamma _{Coll}/\Delta E$ with $\Delta E$ the
average spacing between two states $\mu $, we see that the collective
contribution to $\Gamma _{\mu }$ is proportional to $\Gamma ^{\uparrow
}/N=\Gamma ^{\uparrow }\Delta E/\Gamma _{Coll}$ which is smaller than $%
\Gamma ^{\uparrow }$.

\smallskip

Then one may ask how it is possible to get a large direct decay probability
for the collective state out of small individual decay rates $\Gamma _{\mu }$? 
This is due to a coherence effect. Indeed, if we excite the collective
state $\left| Coll\right\rangle $ and we want to compute its decay
probability, in terms of the individual probability $\Gamma _{\mu }$, using $%
\left| Coll\right\rangle =c_{Coll,\mu }\left| \mu \right\rangle $, we get 
\begin{equation}
\Gamma ^{\uparrow }\propto \left| \left\langle k\right| \Delta \hat{V}\left|
Coll\right\rangle \right| ^{2}=\left| \sum_{\mu }c_{Coll,\mu }\;\left\langle
k\right| \Delta \hat{V}\left| \mu \right\rangle \right| ^{2}
\end{equation}
This coherent sum can be much bigger than the incoherent analogue to $%
\sum_{\mu }\left| c_{Coll,\mu }\right| ^{2}\;\Gamma _{\mu }$. This means
that it is possible to have a strong direct coupling from the collective
state to the continuum built out of a coherent sum of many small decay
probability.

In the following, the direct decay of the collective state is included in
the individual width $\Gamma _\mu $ of each eigenstates $\left| \mu
\right\rangle $. To do so, the coherence should be kept and the width should
fulfill equation (\ref{eq:width}). However, we will often neglect the energy
dependence of the width $\Gamma _\mu $ in the analytical expressions. We
have tested numerically that this energy dependence does not affect the
presented conclusions.

\subsubsection{Microscopic Description of Ericson Fluctuation in Resonant
Phenomena}

Let us first recall the standard description \cite{Boh70} of a doorway state
which includes only the first level. In such a case, assuming $\left|
\left\langle Coll\right| \hat{V}_{1}\left| i_{1}\right\rangle \right| =v_{1},
$ and a regular spectrum of states $\left\{ \left| i_{1}\right\rangle
\right\} $ with a level spacing $E_{i_{1}}-E_{i_{1}-1}=\Delta E_{1}$, the
overlap matrix becomes (see appendix 2) 
\begin{equation}
O_{\mu _{1}}=\frac{v_{1}^{2}}{\left( v_{1}^{2}+v_{1}^{4}\pi ^{2}/\Delta %
E_{1}^{2}\right) +\left( E_{\mu _{1}}-E_{Coll}\right) ^{2}}  \label{usual}
\end{equation}
which is the standard Lorentzian shape with a typical width $\Gamma
_{Coll}^{2}=4\left( v_{1}^{2}+v_{1}^{4}\pi ^{2}/\Delta E_{1}^{2}\right) $.
In the limit of a continuous spectrum and of $\Delta E_{1}\rightarrow 0$
assuming $v_{1}^{2}/\Delta E_{1}=cte$, one gets 
$\Gamma _{Coll}=2\pi v_{1}^{2}/\Delta E_{1}$ which is equivalent to 
the standard Fermi-golden rule.

In this derivation no fluctuation in the strength function is introduced.
This can be traced back to the simplifying assumptions about the density of
states $\left| i\right\rangle $ and the constant interaction matrix elements 
$\left\langle Coll\right| \hat{V}_{1}\left| i_{1}\right\rangle $. Let us,
for example, introduce a fluctuating part of the residual interaction matrix
element defined by $\left| \left\langle Coll\right| \hat{V}_{1}+\delta \hat{V%
}_{1}\left| i_{1}\right\rangle \right| ^{2}=\overline{v}_{1}^{2}(1+\delta
v_{i_{1}})$. Then the overlap matrix can also be splitted into an averaged and
a fluctuating part $O_{\mu _{1}}=\bar{O}_{\mu _{1}}+\delta O_{\mu _{1}}$ ,
the averaged being identical to the constant interaction case (see equation (%
\ref{usual}) and Appendix 3). Introducing diagonal correlation of the
fluctuating matrix element $\delta v_{i_{1}}\delta v_{j_{1}}=c_{i_{1}}\delta
_{i_{1}j_{1}}$, we get 
\begin{equation}
\delta O_{\mu _{1}}\delta O_{\nu _{1}}\simeq \delta _{\mu _{1}\nu
_{1}}\;\delta O_{\mu _{1}}^{2}
\end{equation}
where the expression of $\delta O_{\mu _{1}}^{2}$ is given in appendix 3.
Therefore, one get back the conditions needed to observe Ericson
fluctuations.

\begin{figure}[bh]
\begin{center}
\includegraphics*[width=8cm]{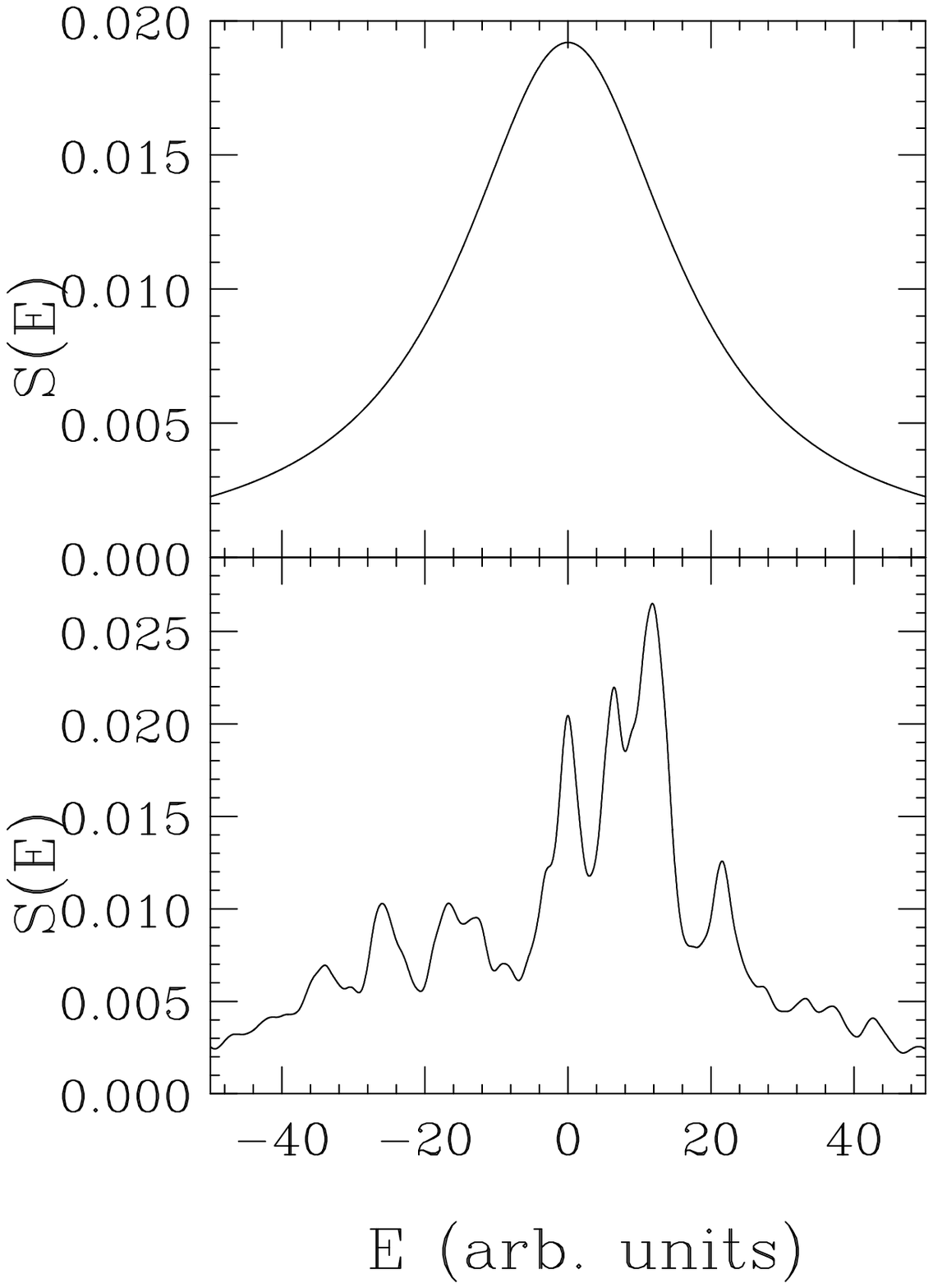}
\end{center}  
\caption{Illustration of the strength computed for a collective doorway
state coupled to a series of decay states damped with a characteristic width
larger than the spacing. Top: when the coupling matrix elements are not
fluctuating. Bottom: when the residual interaction contains random terms. In
both cases, the total width is roughly equal to 30 whereas the individual
width and spacing are $\Gamma _{\mu _{1}}=3$, $\Delta E_{1}=0.5.$ }
\label{fig:2}
\end{figure}

We have investigated this effect using a numerical diagonalization of a
non-fluctuating Hamiltonian $\hat{H}_{1}$ (top part of fig. (\ref{fig:2}))
and of a fluctuating one (bottom part of fig. (\ref{fig:2})). To describe
the finite lifetime of the states $\left| \mu _{1}\right\rangle $, an
imaginary part $-i\;\Gamma _{\mu _{1}}/2$ has been added to each energy $%
E_{\mu _{1}}.$ 
On figure (\ref{fig:2}), we can see that a fluctuating residual interaction
produces Ericson fluctuations. The above discussion applies to the second RPA%
\cite{Dro90,Suh61,Saw62,Dap65,Yan83,Yan86} description of collective excitations
which takes into account the decay of collective states excited through one
body operators, i.e. the coupling of states built from particle-hole (p-h)
type of excitations, into more complex configurations containing
two-particle two-hole (2p-2h) states. In this framework, one would expect to
observe fluctuations of the collective strength related to the
characteristic lifetime of the (2p-2h) states.

\subsubsection{\protect\smallskip Fluctuations within a Microscopic
description of the lifetime of the decay channels}

In order to get a deeper insight into the fluctuation mechanism, one can
introduce the second level of complexity $\left| i_{2}\right\rangle $ . The
corresponding collective response can be computed by introducing as follows:

\begin{itemize}
\item  first the state $\left| Coll\right\rangle $ is coupled to the $\left|
i_1\right\rangle $ by the residual interaction $\hat{V}_{1}+\delta \hat{V}_{1}$
leading to eigenstates $\left| \mu_1 \right\rangle $ 
\begin{equation}
\left| \mu_1 \right\rangle =c_{\mu_1 ,Coll}^{1}\left| Coll\right\rangle
+\sum_{i_1}c_{\mu_1 ,i_1}^{1}\left| i_1\right\rangle 
\end{equation}

\item  Then, the states $\left| \mu_1 \right\rangle $ are coupled through the
residual interaction $\hat{V}_{2}$ with the states $\left| i_2\right\rangle $
in order to built more complex eigenstates $\left| \mu_2 \right\rangle $ 
\begin{equation}
\left| \mu_2 \right\rangle =c_{\mu_2 ,\mu_1 }^{2}\left| \mu_1 \right\rangle
+\sum_{i_2}c_{\mu_1 ,i_2}^{2}\left| i_2\right\rangle 
\end{equation}
\end{itemize}

Therefore we get 
\begin{equation}
\left| \mu_2 \right\rangle =c_{\mu_2 ,\mu_1 }^{2}c_{\mu_1 ,Coll}^{1}\left|
Coll\right\rangle +c_{\mu_2 ,\mu_1 }^{2}\sum_{i_1}c_{\mu_1 ,i_1}^{1}\left|
i_1\right\rangle +\sum_{i_2}c_{\mu_2 ,i_2}^{2}\left| i_2\right\rangle 
\end{equation}
If we assume that the states $\left| \mu _{1}\right\rangle $ are
independently coupled to different ensembles of states $\left|
i_{2}\right\rangle $ through a constant interaction $v_{2}$, then, each $%
\left| \mu _{1}\right\rangle $ will act as a doorway state toward its own
decay channels and so will be spread over a typical width $\Gamma _{\mu
_{1}}^{2}=4\left( v_{2}^{2}+v_{2}^{4}\pi ^{2}/\Delta E_{2}^{2}.\right) $
where $\Delta E_{2}$ is nothing but the level spacing at the states $\left|
i_{2}\right\rangle $ coupled to $\left| \mu _{1}\right\rangle $ . With this
formalism it is clear that the overlap matrix presents in fact two scales 
\begin{equation}
O_{\mu _{2}}=O_{\mu _{1}}^{1}O_{\mu _{2},\mu _{1}}^{2}
\end{equation}
the first one $O_{\mu _{1}}^{1}$ associated with the first level of
complexity $\mu _{1},$ and a second one $O_{\mu _{2},\mu _{1}}^{2}$
associated with the spreading of $\mu _{1}$ over $\mu _{2}$. Within the
presented approximation the strength function reads 
\begin{equation}
S\left( E\right) =\frac{1}{2\pi }\sum_{\mu _{1}}\left( \bar{O}_{\mu
_{1}}+\delta O_{\mu _{1}}\right) v_{2}^{2}\sum_{\mu _{2}}\frac{\Gamma _{\mu
_{2}}}{\left( \left( E_{\mu _{2}}-E_{\mu _{1}}\right) ^{2}+\Gamma _{\mu
_{1}}^{2}/4\right) ^{2}\left( \left( E-E_{\mu _{2}}\right) ^{2}+\Gamma _{\mu
_{2}}^{2}/4\right) }
\end{equation}
where we have introduced a width $\Gamma _{\mu _{2}}$ for the states $\mu
_{2}$. Describing $\bar{O}_{\mu _{1}}$ as in eq. (\ref{usual}) and the $%
\delta O_{\mu _{1}}$ as fluctuation, we recover to the case of Ericson
fluctuations as discussed in the previous chapter. 

\subsection{ Multiscale Fluctuations and the Doorway Hierarchy}

In the previous section, we have assumed that the doorway state was directly
coupled to a large ensemble of states with an average intrinsic width. In
such a case, fluctuations with a typical scale, related to this width, are
observed in the response spectrum. We have also considered the possibility
that the damping might be due to a coupling to more complex states. However,
we have not yet investigated the possibility that this second level of
complexity might also be fluctuating.

\subsubsection{Multiscale Ericson Fluctuation}

We expect in general fluctuations at this second step of the decay, then the
strength reads 
\begin{equation}
S\left( E\right) =\frac{1}{2\pi }\sum_{\mu _{1}}\left( \bar{O}_{\mu
_{1}}^{1}+\delta O_{\mu _{1}}^{1}\right) \sum_{\mu }\left( \bar{O}_{\mu
_{2},\mu _{1}}^{2}+\delta O_{\mu _{2},\mu _{1}}^{2}\right) \frac{\Gamma
_{\mu _{2}}}{\left( E-E_{\mu _{2}}\right) ^{2}+\Gamma _{\mu _{2}}^{2}/4}
\end{equation}
Using a lorentzian shape both for $\bar{O}_{\mu _{1}}^{1}$ with a width $%
\Gamma _{Coll}$ and for $\bar{O}_{\mu _{2},\mu _{1}}^{2}$ with a width $%
\Gamma _{\mu _{1}}$, we can see that two scales $\Gamma _{\mu _{1}}$ and $%
\Gamma _{\mu _{2}}$ are present in the fluctuations of the strength function
on top of the spreading of the collective state $\Gamma _{Coll}$. 
This is illustrated in figure \ref{fig:3} (top-right) 
where we have plotted a strength computed with two scales 
for Ericson fluctuations. 

\subsubsection{Fluctuations of a fragmented strength}

\begin{figure}[h]
\begin{center}
\includegraphics*[height=13.44cm,width=14cm]{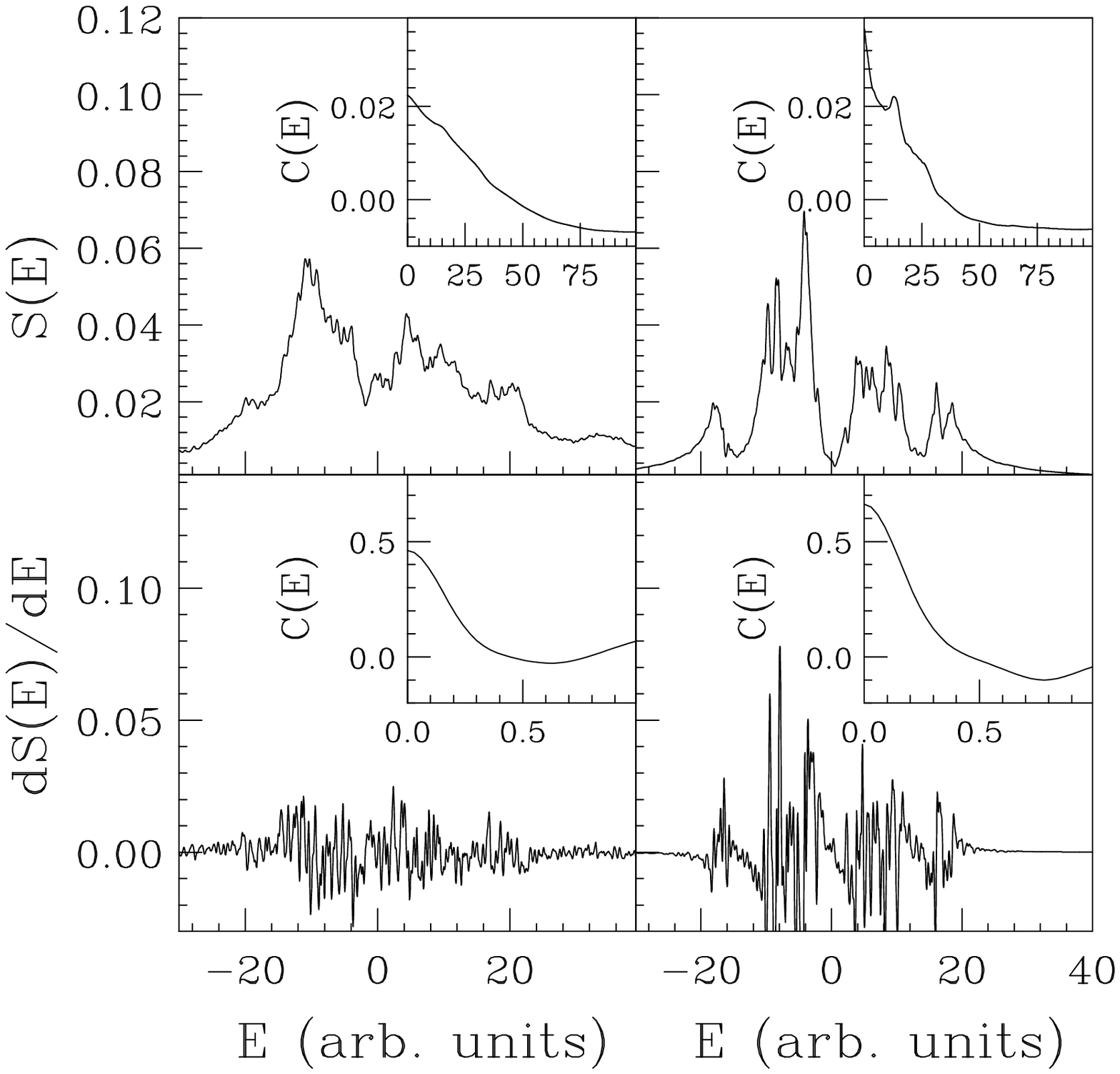}
\end{center}
\caption{Top: Strength distribution presenting multiscale fluctuation. Left:
Ericson fluctuations over two scales(equivalent to fig. 4). Right: Small
Ericson fluctuations on top of a fragmented strength (equivalent to fig. 5).
Bottom: associated derivatives of the strength distributions. In both 
presented cases, the total spreading width is equal to $\Gamma _{Coll}=30$ 
whereas the scales of fluctuations of the two decay channels are 
$\Gamma _{\mu _{1}}=3$ and $ \Gamma _{\mu _{2}}=0.5$. In each figures, 
the associated autocorrelation function respectively performed
on the strength (top part) or on the derivative of the strength (bottom part) 
are presented in insert.}
\label{fig:3}
\end{figure}

Another interesting situation is the case of Landau spreading. Indeed, at
the RPA level of description\cite{Rin81,Row70}, the strength often appears
splitted in several components $\left|Coll_{n}\right>$. In such a case, at
least three scales might be identified. The Landau spreading can be directly
observe as the fragmentation of the strength while the width of each
fragment can be identified with the lifetime of each individual collective
state. Moreover, as discussed above the strength presents fluctuations on
top of each resonant line-shape which are characteristic of the various
decay channels widths. Indeed, if we assume that the different fragments are
decaying toward independent complex states, the strength is simply the sum
of the various components associated with a collective excitation $%
\left|Coll_n\right\rangle $ 
\begin{equation}
S\left( E\right) =\sum_n \left| \left\langle Coll \right. \left|
Coll_n\right\rangle \right| ^2 \left( \sum_\mu \frac{\left| \left\langle \mu
\right. \left| Coll_n\right\rangle \right| ^2}{\left( E-E_\mu \right)
+i\Gamma _\mu /2}\right)
\end{equation}
Such a fragmented strength is illustrated in figure \ref{fig:3} (top-right). 

\section{Observation of fluctuations at different scales}

The observation of the collective state width $\Gamma_{Coll}$, the largest
scale in the strength, is rather straightforward using standard techniques
such as fits, variance estimation or even autocorrelation calculations. The
extraction of characteristic widths from a spectrum presenting many
different scales of fluctuations is a more complicated problem. In this
chapter, we discuss several methods and propose new approaches for
extracting signals of multiscale fluctuations.

\subsection{Standard method and its extension.}

We have seen in the previous chapter that the autocorrelation function is a
useful tool when fluctuations and fragmentation of the response are involved
(see figure (\ref{fig:1})). 
On the one hand, when the autocorrelation techniques is used directly on the
strength distribution, the obtained signal can be related the total width of
the collective mode. A standard technique to extract the fluctuation
properties of a fine structure on top of some smooth ''background'' is to
subtract an averaged distribution\cite{Kil87} from the signal. However, the
obtained result appears to be dependent on the method and parameters (such
as the smoothing interval) used to define this average strength. In order to
overcome this problem, we have performed the autocorrelation analysis on the
derivative of the strength distribution. The derivative is very sensitive to
the fluctuations over the smaller scale present in the spectrum since they
are associated with very rapid variations. In this case, the autocorrelation
function gives half of the width of the smallest fluctuations scale.

These two types of autocorrelation analyses, on the strength and on its
derivative, are illustrated in figures (\ref{fig:3})
for two cases: a multiscale (two scales) Ericson fluctuation (left of fig. 
(\ref{fig:3})) and a fragmented strength with fine structure
(right of fig. (\ref{fig:3})). 

In both cases (multiscale Ericson of fragmented), 
the largest and the smallest scales in the fluctuations can be
extracted. Considering now the fragmented strength case, an intermediate
scale can also be observed. However, when Ericson fluctuations occurs at
many different scales, the autocorrelations techniques applied to the
strength and its derivative only gives access to the largest and the
smallest scales of the fluctuations and intermediates scales could not
be obtained.

During the past years, different alternative techniques have been 
proposed to extract properties of fluctuations in the decay of
collective motions \cite{Dro94,Dro95,Dro96,Ber99,Aib99}. In particular, 
in a second RPA picture, where only two-body correlations are retained, 
a possible fractal nature of the fluctuations has been discussed
\cite{Dro95,Dro96,Ber99,Aib99}. 
When higher correlations are considered, we have shown that a hierarchy
of well-separated scales may also be present. In the next section,
we will propose a novel technique, that may help to get either the
well separated scales or the fractal nature of nuclear decay.

\subsection{Entropy index for multiscale fluctuation}

Recently, a method based on the definition of an entropy index has been
proposed by Hwa\cite{Hwa98} in order to extract scaling behavior in
fluctuating signal. The entropy index appears as a good indicator of the
existence of different scales in a strongly fluctuating spectrum. We have
adapted this method to the nuclear response case.

The total energy interval $\Delta E=E_{MAX}-E_{MIN},$ is first divided into $%
n$ bins of resolution $\delta E$ ($n=\Delta E/\delta E$). The signal in each
bin is analyzed according to a simple overlap with a step function changing
from $-1$ to $+1$ in the middle of the considered bin\footnote{%
This method can be viewed as a wavelet analysis.}. Therefore, at every scale 
$\delta E$, in each of the bin $j$, a coefficient $D_j\left( \delta E\right) 
$ is defined as

\begin{equation}
D_j\left( \delta E\right) =\int_{E_{MIN}+\left( j-1\right) \delta
E}^{E_{MIN}+j\delta E}dE~S\left( E\right) {\rm sign}\left( E-\left(
j-1/2\right) \delta E\right)
\end{equation}
The coefficients $D_j$ can be considered as a coarse grained derivative of $%
S $. In order to focus on the global properties of the fluctuations at a
given size $\delta E$ , an entropy factor $K\left( \delta E\right) $ can be
defined as

\begin{equation}
K\left( \delta E\right) =\frac 1n\sum_{j=1,n}W_j\left( \delta E\right) \log
W_j\left( \delta E\right)
\end{equation}
where the coefficient $W_j\left( \delta E\right) =$ $D_j/\left\langle
D_j\right\rangle $ are nothing but the coefficients $D_j$ normalized to
their averaged value ($\left\langle D_j\right\rangle =1/n\sum_{j=1,n}D_j$).

\begin{figure}[h]
\begin{center}
\includegraphics*[height=15cm]{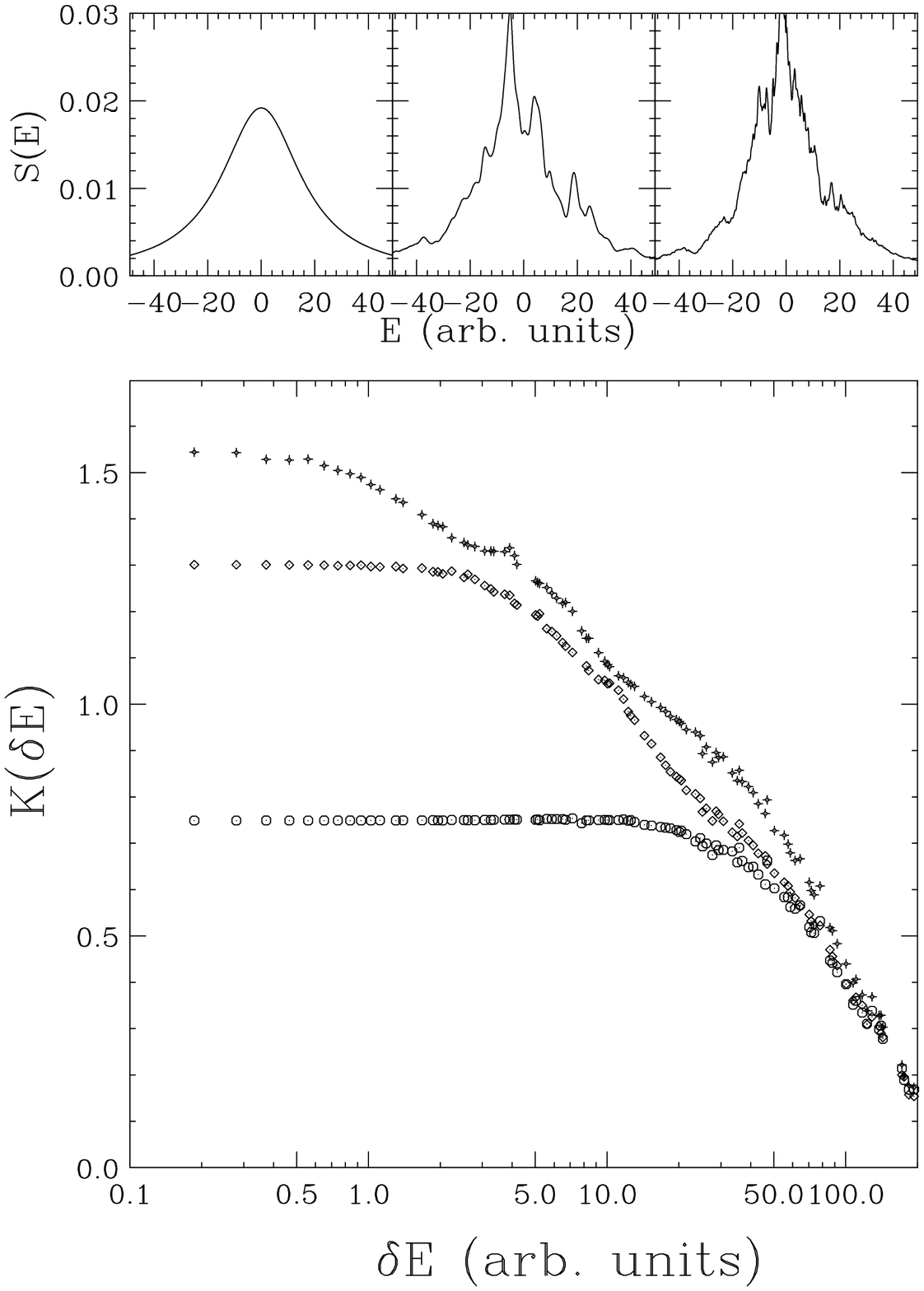}
\end{center}                                       
\caption{Top: from left to right, strength function with respectively one, 
two and three scales of fluctuations ($\Gamma =30$, $3$ and $0.5$) are 
displayed. Bottom: Evolution of $K\left( \delta E\right) $ as a 
function of the bin size $\delta E$ for the three considered strength 
function with one (circle), two (diamond) and three scales (cross).}
\label{fig:5_2}
\end{figure}

\begin{figure}[h]
\begin{center}
\includegraphics*[height=10cm]{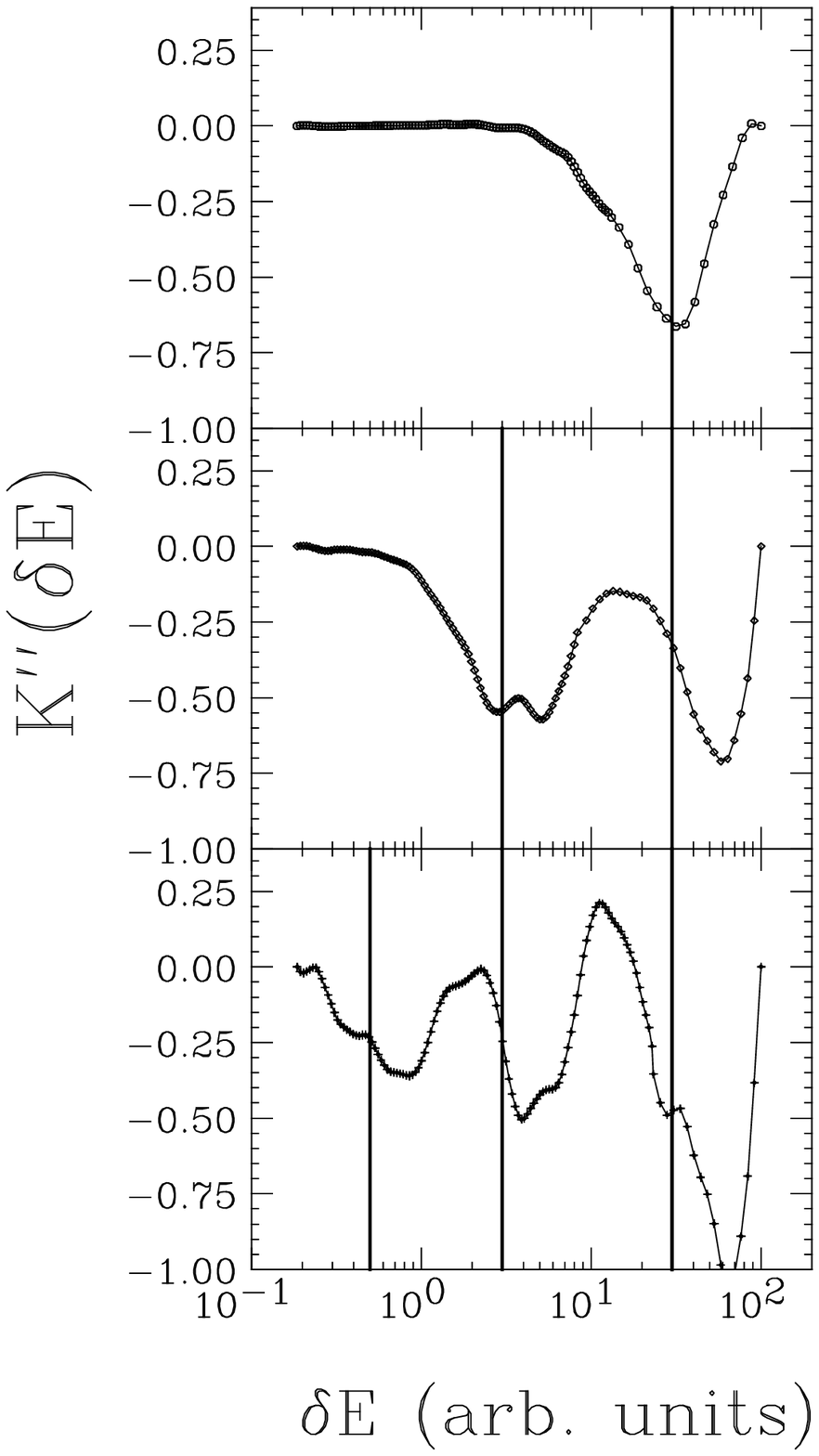}
\end{center}                                       
\caption{Numerical estimation of the second derivative 
(noted $K''\left(\delta E\right)$) of the entropy index as a function of the 
bin size $\delta E$. $K''(\delta E)$ is 
associated with strengths where respectively one (top), two (middle) and three 
(bottom) scales of fluctuations are present. In each figures, the scales of
fluctuations are indicated by thick vertical lines.}
\label{fig:5_der}
\end{figure}

In the reference\cite{Hwa98}, it is shown that a linear decrease of $K\left(
\delta E\right) $ as a function of $\log \left( \delta E\right) $
characterizes the existence of fluctuations at all scales. When fluctuations
with specific scales are considered, a different behavior is expected.
Having in mind that the coefficients $W_j\left( \delta E\right) $ might be
interpreted as a normalized coarse-grained derivative, $K\left( \delta
E\right)$ should remain almost constant between two typical scales since
in-between these scales, the coarse grained derivative of $S$ do not vary
much. However, when going from one scale to another, the derivative looses
part of its structure and one is expecting an entropy variation. 
This situation is illustrated in figure (\ref{fig:5_2}) for three different
cases, where respectively fluctuations over one, two and three scales have
been introduced in the strength. Note that, in order to avoid
problems due to the limited number of bins for large 
bin size, we have considered a function defined as a 
repetition of the strength function instead of the strength itself.
In practice, we used 31 repetitions of the strength.
The entropy index is then applied on the new function inside a 
large energy interval, enabling
to have a large number of bins in the energy region under interest 
($\delta E \le \Delta E$). 
From fig. (\ref{fig:5_2}), one can observe that the evolution of $K\left(
\delta E\right)$ indicates the presence of respectively one, two and 
three scales by a change in its curvature. In order to emphasize this
evolution, we have plotted in figure (\ref{fig:5_der}) a numerical
estimation of the second derivative $K''\left(\delta E\right)$ of the 
entropy index. 
The presence of one, two or three scales is thus signed by the presence
of respectively one, two and three minima in these second derivative.   

In all the presented cases, $K (\delta E) $ is a good indicator
of the different characteristic scales in the nuclear response. Using the
positions of the curvature variations, the various scales can be roughly 
estimated.

In this section, the entropy  index method was applied to 
model where scales are well separated. This situation
may append at rather low excitation energy where few degrees 
of freedom are coupled to collective states\cite{Lac98}. When 
excitation energy increases, the number of internal and external
degrees of freedom involved in the coupling becomes very
large\cite{Fra96} and a
statistical treatment is required \cite{Boh70,Kil87,Sok88}. In this
case, many different scales are expected and can be uncovered through
the entropy index method. This method can also be used in more complex
situations where mixing of the different scales is expected. It can even
sign self-similar fluctuations \cite{Dro94,Dro95,Ber99}. 
It should be noticed that the smallest scales expected  
in the strength are related to the lifetime of the compound nucleus
states which are less than the eV. However, experimental resolutions
have not yet reached this degree of accuracy \cite{Kam97,Ott98}.
Elaborated  techniques based on statistical assumptions have been
applied in order to extract the properties of invisible fine
structures\cite{Kil87}. The entropy index method can be viewed as a
model independent way  to extract information about fluctuations. Only
scales larger than the experimental resolution can be accessed but no
assumptions on the statistical properties of the studied spectrum are
needed.

\section{ Critical discussion of the Doorway hierarchy picture}

In order to observe multiscale fluctuations, we have assumed a particular
hierarchy in the Hamiltonian following the general scenario of a gradual
complexification of a collective motion until it reaches the compound
nucleus chaotic states. It is in fact the implicit assumption of many
simulations such as the extended mean-field approach which takes into
account the two-body collisions as a damping mechanism\cite
{Lac98,Won79,Bal86,Abe96,Ayi85}. In these cases, a
hierarchy of couplings is implicitly assumed, the main ansatz being that
cutting this hierarchy at any level leads to an approximate strength which
is only slightly modified when the next level is introduced. From a quantal
point of view, this means that an ensemble of states ordered by increasing
complexity, such as states with increasing number of quasi-particle
excitations, can be defined and that each level of complexity can be
considered as a perturbation on top of the previous level. However, there is
no {\it a priori} reason that such a robust hierarchy exists. Indeed, in
quantum mechanics, a modification of few matrix elements is often sufficient
to introduce an important rearrangement of the whole spectrum. Therefore,
even if the hierarchy of doorway seems valid for several decay steps, it is
always possible that the introduction of the next level of complexity deeply
transforms the overall picture.
For example, the coupling of overlapping resonances through their common 
 decay channels has been already discussed in connection with
nuclear relaxation  (see for
instance \cite{Mah69,Sok97-2}). Specific coherent effects, like the Dicke 
supperradiance well known in quantum optic\cite{Dic54}, could also be
present in nuclear spectra and  could suppress Ericson
fluctuations\cite{Sok89}.  Such coherence effects can also 
affect the nuclear response when
a hierarchy of intrinsic degrees of freedom is considered.

Let us consider a single collective doorway state $\left| Coll\right\rangle $
damped through a residual interaction $\hat{V}_{1}$ and $\hat{V}_{2}$
assuming that $\left\langle Coll\right| \hat{V}_{1}\left| i_{1}\right\rangle
=v_{1},$ $\left\langle i_{1}\right| \hat{V}_{2}\left| i_{2}\right\rangle
=v_{2}$ and that the second level of complexity $\left| i_{2}\right\rangle $
is characterized by a level spacing $\Delta E_{2}$. Then we can take the
limit $v_{2}^{2}\rightarrow 0$ and $\Delta E_{2}\rightarrow 0$ keeping $%
\Gamma _{1}=2\pi v_{2}^{2}/\Delta E_{2}$ constant$.$ In the case of a
regular ensemble of states $i_{1}$ with a spacing $\Delta E_{1}$, we can
write the overlap matrix (see Appendix 4) 
\begin{equation}
O_{\mu _{2}}=\left| c_{Coll}^{\mu _{2}}\right| ^{2}=\frac{v_{2}^{2}v_{1}^{2}%
}{\frac{\Gamma _{1}^{2}}{4}\left( E_{\mu _{2}}-E_{Coll}\right) ^{2}+\frac{^{%
\Delta E_{1}^{2}}}{\pi ^{2}}\left( \left( E_{\mu _{2}}-E_{Coll}\right) \tan
\left( \pi \frac{E_{\mu _{2}}}{\Delta E_{1}}\right) -\frac{\Gamma _{Coll}}{2}%
\right) ^{2}}  \label{O123}
\end{equation}
where we have introduced $\Gamma _{Coll}=2\pi v_{1}^{2}/\Delta E_{1}$. When $%
\Gamma _{1}\gg \Delta E_{1}$. In such a case, the overlap $O_{\mu _{2}}$ has
a typical scale of fluctuation not related to $\Gamma _{1}$ but to $\Delta %
E_{1}$ since the denominator goes to infinity each $E_{\mu _{2}}=(n+1/2)%
\Delta E_{1}$ in eq. (\ref{O123}).

\begin{figure}[h]
\begin{center}
\includegraphics*[height=7.8cm,width=10cm]{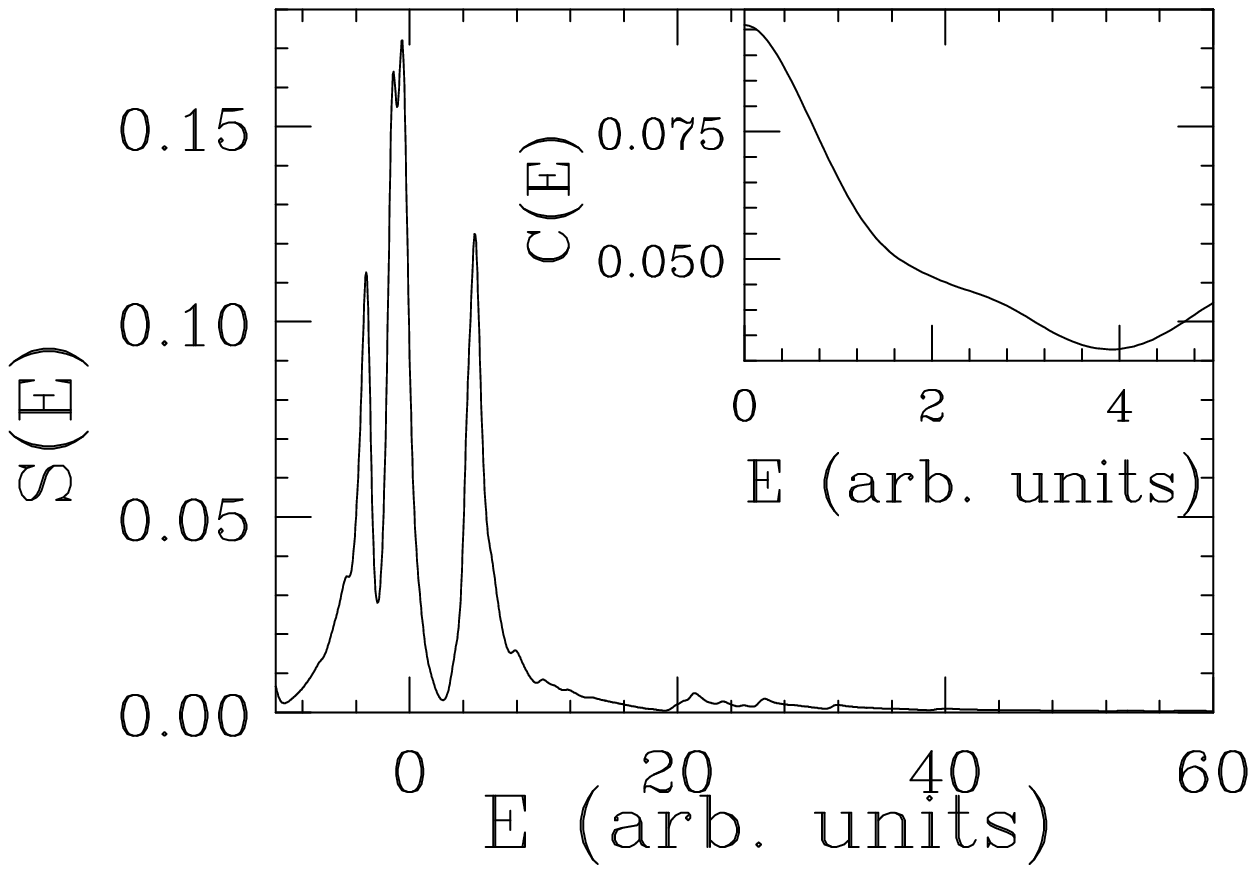}
\end{center}
\caption{Illustration of the strength computed for a doorway state coupled
to many decay states themselves coupled to an ensemble of states and
computed introducing all types of fluctuations in the residual hamiltonian.
For the calculation, we have taken $\Gamma _{Coll}=100,$ $\Delta E_{1}=3$ , $%
\Gamma _{1}=30$ and $\Delta E_{2}=0.5.$ In addition, a small width $\Gamma
_{\mu _{2}}=1$ has been assumed for the eigenstates of the hamiltonian in
order to draw the strength function. In insert, the autocorrelation function
is plotted demonstrating that the widths are proportional to $\Delta E_{1}$. 
}
\label{fig:11}
\end{figure}

Again we have tested that all type of fluctuations of the residual
Hamiltonian as illustrated in figure (\ref{fig:11}). This figure presents
strong oscillations with a width related to $\Delta E_1 $ as shown by the
autocorrelation function. Therefore, the occurrence of interference like
pattern characteristic of the level spacing $\Delta E_1$ should be
considered as generic. We would also like to mention that a strong coupling
between the first and second decay channel induces a narrowing of the
apparent width of the collective states. This effect is analogous to the
motional narrowing described in\cite{Bro87}. 

This illustration stresses the fact that in the case of a coupling of
several states through their decay channels one should expect interference
pattern due to a feedback of the decay channels on the properties and the
coupling of the doorway states. Then, the fluctuation pattern can reflect
the typical level spacing and not the typical width.

\section{Conclusion}

In conclusion, we have studied the role of several steps in the decay of
doorway states 
%both from the analytical and also from the numerical point of
%view. 
We have shown that, in some cases, in particular when the decay channels of
each states encountered at each level of the decay cascade can be considered
as independent, one should expect to observe several scales in the
fluctuation pattern of the response function. In such a case, the various
fluctuations are characteristic of the typical lifetime at each level of
complexity encountered in the path from the doorway toward the chaotic
compound nucleus states. We have shown that this multiscale Ericson
fluctuations can be partially revealed by applying the autocorrelation
technique on the strength. However, standard methods which involve different
smoothing procedures seem to bias the obtained results. In order to overcome
this problem, we have proposed two new techniques: the first one consists in
computing the autocorrelation function on the derivative of the strength
distribution in order to extract the smallest scale in the fluctuations, the
second one, called Entropy index, gives information on how many scales does
exist in the fluctuation pattern.Therefore, high resolution experiments are
indeed very interesting tools to study the damping mechanism of collective
states.

Finally, the analytical results as well as the numerical simulations of the
last chapter illustrate that a hierarchy of scales in the fluctuation,
directly related to the lifetime of each level of complexity, is not the
generic case and that often the fluctuations can be related to other
physical quantities. In particular we have investigated the effects of decay
channels interacting via the states encountered at a higher level of
complexity. In this case, interferences between different decay channels are
found and the induced fluctuations are not related to the decay width of the
considered states but to their spacing. In such a case, experiment with a
very high resolution will present fluctuations which are characteristic of
the level spacing and not of the width as in Ericson fluctuations.

Caution should thus be used when interpreting 
%extracting the information from 
the observed width of the fluctuations. Indeed, the interferences, due to an
interaction of some states sharing the same decay channels, may produce
narrow fluctuations which may mimic long lived systems. This is an important
finding which can modify the interpretation of Experimental data.

\smallskip

{\LARGE Acknowledgments:}

\smallskip We want to thank the experimental group working on giant
resonances and in particular S. Ottini, W. Mittig, P. Roussel-Chomaz and A.
Drouart, who have attracted our attention to this problem. This article has
benefit from many discussions with them.

\appendix{\bf APPENDIX 1: Ericson Fluctuations in doorway processes}

\smallskip

Let us consider the spectral response  
\begin{equation}
R\left( E\right) =\sum_\mu \frac{\left| \left\langle \mu \right. \left|
Coll\right\rangle \right| ^2}{\left( E-E_\mu \right) +i\Gamma _\mu /2}
\end{equation}

When we introduce an averaged overlap and a fluctuating one, $\left|
\left\langle \mu \right. \left| Coll\right\rangle \right| ^2 =\bar{O}\left(
E_\mu \right) + \delta O_\mu$, the response can also be split into two
components. On the one hand, it contains an averaged response function $\bar{%
R}\left( E\right) =$ $\int_{\delta E}dE^{\prime }\;R\left( E^{\prime
}\right) $ which can be written as for $E>0$ 
\begin{equation}
\bar{R}\left( E\right) \simeq \frac 1{\delta E}\int_{\delta
E}dE^{\prime}\;\rho \left( E^{\prime}\right) \;\frac{\bar{O}\left(
E^{\prime}\right) }{\left( E-E^{\prime}\right) +i\overline{\Gamma }/2}\simeq 
\bar{\rho}\;\bar{O}\left( E\right)
\end{equation}
where we have introduced the density of states $\rho \left( E\right)
=\sum_\mu \delta \left( E-E_\mu \right) $, the average density of states $%
\bar{\rho}=\int_{\delta E}dE^{\prime}\;\rho \left( E^{\prime}\right) /\delta
E=1/\Delta E$ and an averaged width $\overline{\Gamma }$ (the integration
interval $\delta E$ being taken much larger than $\overline{\Gamma }$). On
the other hand remains a fluctuating term 
\begin{equation}
\delta R\left( E\right) =\sum_\mu \frac{\delta O_\mu }{\left( E-E_\mu
\right) +i\Gamma _\mu /2}
\end{equation}
These fluctuations still contain information about the averaged width $%
\overline{\Gamma }$ as it was demonstrated by Ericson\cite{Eri63}. Indeed,
the autocorrelation function of this noise is given by 
\begin{eqnarray}
C_R(E) &=&\frac 1{\delta E}\int_{E_0}^{E_0+\delta E}dE^{\prime }\;\delta
R^{*}\left( E^{\prime }\right) \;\delta R\left( E^{\prime }+E\right) \\
C_R(E) &=&\sum_{\mu \nu }\frac 1{\delta E}\int_{E_0}^{E_0+\delta
E}dE^{\prime }\frac{\delta O_\mu }{E^{\prime }-{\cal E}_\mu ^{*}}~\ \frac{%
\delta O_\nu }{E^{\prime }+E-{\cal E}_\nu }
\end{eqnarray}
where we have introduced complex energy notations ${\cal E}_\mu =E_\mu
-i\Gamma _\mu /2.$ If we assume that $\delta E\gg \overline{\Gamma }$, we
can extend the integration to infinity, so that, taking advantage of the
relation $\left( x-y\right) ^{-1}\left( x-z\right) ^{-1}=\left( y-z\right)
^{-1}\left( \left( x-y\right) ^{-1}-\left( x-z\right) ^{-1}\right) $, it is
possible to perform a Cauchy integral leading to 
\begin{equation}
C_R(E)=\sum_{\mu \in \Delta E}\frac{2i\pi }{\delta E}\sum_\nu \frac{\delta
O_\mu \delta O_\nu }{{\cal E}_\nu -E-{\cal E}_\mu ^{*}}+\mu
\longleftrightarrow \nu
\end{equation}
If we assume random correlations between the overlap fluctuation 
\begin{equation}
\delta O_\mu \delta O_\nu =\delta O_\mu ^2\;\delta _{\mu \nu }
\label{eq:correl}
\end{equation}
in average, we get the correlation 
\begin{equation}
C_R(E)=\frac{4\pi }{\Delta E}\frac{\overline{\delta O^2}}{E^2+\overline{%
\Gamma }^2}
\end{equation}
A similar equation holds for the strength function associated to equation (%
\ref{resp}) 
\begin{equation}
S\left( E\right) =-\frac 1\pi Im\left( R(E)\right) =\sum_\mu \frac{\frac{%
\Gamma _\mu }{2\pi }\left| \left\langle \mu \right. \left| Coll\right\rangle
\right| ^2}{\left( E-E_\mu \right) ^2+\Gamma _\mu ^2/4}
\end{equation}
Indeed, we can define the autocorrelation function of the strength
distribution as 
\begin{equation}
C(E)=\frac 1{\delta E}\int_{E_0}^{E_0+\delta E}dE^{\prime }\;\delta S\left(
E^{\prime }\right) \;\delta S\left( E^{\prime }+E\right)
\end{equation}
Noting that $\displaystyle S(E)=-1/(2i\pi )\left( R(E)-R^{*}(E)\right) $ we
have 
\begin{equation}
C(E)=\frac 2{\pi \Delta E}\frac{\overline{\delta O^2}}{E^2+\overline{\Gamma }%
^2}
\end{equation}
This relation shows that, even in the case of strongly overlapping states,
one can still extract information about the typical width by looking at the
fluctuating part of the considered autocorrelation spectrum as it was
demonstrated in \cite{Eri63}.

\smallskip

\appendix{\bf APPENDIX 2: Simple Doorway Picture of a Damping Mechanism}

\smallskip

The diagonalization of the Hamiltonian $\hat{H}_{1}=\hat{H}_{0}+\hat{V}_{1}$
produces eigenstates $\left| \mu _{1}\right\rangle $ associated with the
eigenenergies $E_{\mu _{1}}$ which fulfill the following dispersion relation 
\begin{equation}
E_{\mu _{1}}-E_{Coll}=f_{1}\left( E_{\mu _{1}}\right) v_{1}^{2}
\end{equation}
where we have considered a constant interaction $v_{1}^{2}=\left|
\left\langle Coll\right| \hat{V}\left| i_{1}\right\rangle \right| ^{2}$. The
overlap matrix $O_{\mu _{1}}=\left| \left\langle \mu _{1}\right. \left|
Coll\right\rangle \right| ^{2}$ reads 
\begin{equation}
O_{\mu _{1}}=\frac{1}{1+f_{2}\left( E_{\mu _{1}}\right) v_{1}^{2}}
\end{equation}
In the previous equations, we have introduced the functions $f_{n}$%
\begin{equation}
f_{n}\left( E\right) =\sum_{i_{1}}\frac{1}{\left( E-E_{i_{1}}\right) ^{n}}
\end{equation}
which are related by a recurrence relation 
\begin{equation}
f_{n+1}\left( E\right) =-n\frac{\partial f_{n}\left( E\right) }{\partial E}
\label{eq:g_f}
\end{equation}

In order to get the usual Lorentzian shape, we assume a regular spectrum of
energy $E_{i_{1}}$: $E_{i_{1}}-E_{i_{1}-1}=\Delta E_{1}$ with $i_{1}$
running over both positive and negative integer values. Then, using the
relation $\sum_{i}1/\left( x-i\right) =\pi \cot \left( \pi x\right) ,$ the
dispersion relation becomes 
\begin{equation}
E_{\mu _{1}}-E_{Coll}=f_{1}\left( E_{\mu _{1}}\right) =\frac{v_{1}^{2}\pi }{%
\Delta E_{1}}\cot \left( \pi \frac{\left( E_{\mu _{1}}-E_{Coll}\right) }{%
\Delta E_{1}}\right)   \label{eq:disp}
\end{equation}
On the other hand, using the relation (\ref{eq:g_f}) and taking advantage of
the dispersion relation (\ref{eq:disp}), $f_{2}\left( E_{\mu _{1}}\right) $
can be recast as 
\begin{equation}
f_{2}\left( E_{\mu _{1}}\right) =\left( \frac{\pi ^{2}}{\Delta E_{1}^{2}}+%
\frac{\left( E_{\mu _{1}}-E_{Coll}\right) ^{2}}{v_{1}^{4}}\right) 
\end{equation}
so that the overlap matrix simply reads 
\begin{equation}
O_{\mu _{1}}=\frac{v_{1}^{2}}{\left( v_{1}^{2}+v_{1}^{4}\pi ^{2}/\Delta %
E_{1}^{2}\right) +\left( E_{\mu _{1}}-E_{Coll}\right) ^{2}}
\end{equation}
which is the standard Lorentzian shape with a typical width $\Gamma
_{Coll}^{2}=4\left( v_{1}^{2}+v_{1}^{4}\pi ^{2}/\Delta E_{1}^{2}\right) $.

\smallskip

\appendix{\bf APPENDIX 3: Microscopic Description of Ericson Fluctuation in
Resonant Phenomena}

\smallskip

Following Appendix 2, let us, for example, introduce a small fluctuating
matrix $\delta \hat{V}_{1}.$ Then we define a fluctuating part for the $f_{n}
$ functions 
\begin{equation}
\delta f_{n}\left( E\right) =\sum_{i_{1}}\frac{\delta v_{i_{1}}}{\left(
E-E_{i_{1}}\right) ^{n}}
\end{equation}
where we have used the notation 
%introduced the fluctuating part of the residual interaction
%matrix element defined by 
$\left| \left\langle Coll\right| \hat{V}_{1}+\delta \hat{V}_{1}\left|
i_{1}\right\rangle \right| ^{2}=\overline{v_{1}}^{2}(1+\delta v_{i_{1}})$ .
The new dispersion relation reads 
\begin{equation}
E_{\mu _{1}}-E_{Coll}=\overline{v_{1}}^{2}(f_{1}\left( E_{\mu _{1}}\right)
+\delta f_{1}\left( E_{\mu _{1}}\right) )
\end{equation}
while the overlap becomes 
\begin{equation}
O_{\mu _{1}}=\frac{1}{1+\overline{v_{1}}^{2}\left( f_{2}\left( E_{\mu
_{1}}\right) +\delta f_{2}\left( E_{\mu _{1}}\right) \right) }
\end{equation}
In the case of a regular spectrum of states $\left| i_{1}\right\rangle ,$
using the relations derived above, the dispersion relation becomes 
\begin{equation}
E_{\mu _{1}}-E_{Coll}=\overline{v_{1}}^{2}\left( \frac{\pi }{\Delta E_{1}}%
\cot \left( \pi \frac{\left( E_{\mu _{1}}-E_{Coll}\right) }{\Delta E_{1}}%
\right) +\delta f_{1}\left( E_{\mu _{1}}\right) \right) 
\end{equation}
so that, we may use this relation in order to compute $f_{2}\left( E_{\mu
_{1}}\right) $%
\begin{equation}
f_{2}\left( E_{\mu _{1}}\right) =\left( \frac{\pi ^{2}}{\Delta E_{1}^{2}}+%
\bar{v_1}^{-4}\left( E_{\mu _{1}}-E_{Coll}-\overline{v_{1}}^{2}\delta
f_{1}\left( E_{\mu _{1}}\right) \right) ^{2}\right) 
\end{equation}
The latter equation can be introduced in the overlap matrix leading the
perturbative result 
\begin{equation}
O_{\mu _{1}}=\bar{O}_{\mu _{1}}+\delta O_{\mu _{1}}
\end{equation}
with an averaged value, $\bar{O}_{\mu _{1}},$ identical to the one derived
in appendix 2 and with a fluctuating part given by 
\begin{equation}
\delta O_{\mu _{1}}=\bar{O}_{\mu _{1}}^{2}\left( 2\left( E_{\mu
_{1}}-E_{Coll}\right) \delta f_{1}\left( E_{\mu _{1}}\right) -\overline{v_{1}%
}^{2}\delta f_{2}\left( E_{\mu _{1}}\right) \right) 
\end{equation}
We finally get : 
\begin{equation}
\delta O_{\mu _{1}}=\bar{O}_{\mu _{1}}^{2}\sum_{i_{1}}\delta v_{i_{1}}\delta
o_{\mu _{1}}\left( E_{i_{1}}\right) 
\end{equation}
with
\begin{equation}
\delta o_{\mu _{1}}\left( E\right) =\frac{1}{\left( E_{\mu _{1}}-E\right) }%
\left( \frac{2\left( E_{\mu _{1}}-E_{Coll}\right) }{\bar{v_1}^{2}}-\frac{1}{%
E_{\mu _{1}}-E}\right) 
\end{equation}
This is a fluctuating correction to the averaged overlap which looks like
the one needed in order to observe Ericson-like fluctuations. Indeed, the
correlation $\overline{\delta O_{\mu _{1}}\delta O_{\nu _{1}}}$ reads as a
function of the correlation of the fluctuating matrix element $\overline{%
\delta v_{i_{1}}\delta v_{j_{1}}}=c_{i_{1}j_{1}}$%
\begin{equation}
\overline{\delta O_{\mu _{1}}\delta O_{\nu _{1}}}=\bar{O}_{\mu _{1}}^{2}\bar{%
O}_{\nu _{1}}^{2}\sum_{i_{1}j_{1}}c_{i_{1}j_{1}}\delta o_{\mu _{1}}\left(
E_{i_{1}}\right) \delta o_{\nu _{1}}\left( E_{j_{1}}\right) 
\end{equation}
If we now assume uncorrelated fluctuations for the matrix elements $%
c_{i_{1}j_{1}}=c_{i_{1}}\delta _{i_{1}j_{1}}$ we get 
\begin{equation}
\overline{\delta O_{\mu _{1}}\delta O_{\nu _{1}}}=\bar{O}_{\mu _{1}}^{2}\bar{%
O}_{\nu _{1}}^{2}\sum_{i_{1}}c_{i_{1}}\delta o_{\mu _{1}}\left(
E_{i_{1}}\right) \delta o_{\nu _{1}}\left( E_{i_{1}}\right) 
\end{equation}
because of the dispersion relation, the energy $E_{\mu _{1}}$ is always very
close from one particular state $\left| i_{1}\right\rangle $ which we note $%
\left| i_{1}^{\mu _{1}}\right\rangle $ associated to an unperturbed energy $%
E_{i_{1}^{\mu _{1}}}.$ Then, in $\delta o_{\mu _{1}}$ $,$ mainly this state $%
i_{1}^{\mu _{1}}$ is participating to the sum over $i_{1}$ so that $\delta
O_{\mu _{1}}\approx \bar{O}_{\mu _{1}}^{2}\delta v_{i_{1}^{\mu _{1}}}\bar{o}%
_{\mu _{1}}\left( E_{i_{1}^{\mu _{1}}}\right) $ and the correlation reads 
\begin{equation}
\overline{\delta O_{\mu _{1}}\delta O_{\nu _{1}}}\approx \delta _{\mu
_{1}\nu _{1}}\;\bar{O}_{\mu _{1}}^{4}\;c_{i_{1}^{\mu _{1}}}\delta o_{\mu
_{1}}^{2}\left( E_{i_{1}^{\mu _{1}}}\right) =\delta _{\mu _{1}\nu
_{1}}\;\delta O_{\mu _{1}}^{2}
\end{equation}
Therefore, one get back the correlations needed to observe Ericson
fluctuations. However, it should be noticed that compared with the simple
constant correlation assumed in eq. (\ref{eq:correl}) one may have here a
smooth energy dependence of $\delta O_{\mu _{1}}^{2}.$ This however do
affect the conclusions reached about the width of the autocorrelation
function.

\appendix{\bf APPENDIX 4: Interferences in the Damping of a Resonance}%
\smallskip 

\smallskip

We can finally consider the related case of a single doorway collective
state $\left| Coll\right\rangle $ damped through a residual interaction $%
\hat{V}_{1}$ toward a first level of complexity described by the states $%
\left| i_{1}\right\rangle $ which are then coupled to many states $\left|
i_{2}\right\rangle $ by a second part of the residual interaction $\hat{V}%
_{2}$. Then, the diagonalization of $\hat{H}=\hat{H}_{0}+\hat{V}_{1}+\hat{V}%
_{2}$ (where $\left| Coll\right\rangle $, $\left| i_{1}\right\rangle $ and $%
\left| i_{2}\right\rangle $ are eigenstates of $\hat{H}_{0}$) produces
eigenstates 
\begin{equation}
\left| \mu _{2}\right\rangle =c_{\mu _{2},Coll}^{{}}\left| Coll\right\rangle
+\sum_{i_{1}}c_{\mu _{2},i_{1}}^{{}}\left| i_{1}\right\rangle
+\sum_{i_{2}}c_{\mu _{2},i_{2}}^{{}}\left| i_{2}\right\rangle 
\end{equation}
The Schr\"{o}dinger equation leads to 
\begin{eqnarray}
c_{\mu _{2},Coll}^{{}}\left( E_{\mu _{2}}-E_{Coll}\right) 
&=&\sum_{i_{1}}c_{\mu _{2},i_{1}}^{{}}\left\langle Coll\right| \hat{V}%
_{1}\left| i_{1}\right\rangle =v_{1}a_{\mu _{2}}^{1} \\
c_{\mu _{2},i_{1}}^{{}}\left( E_{\mu _{2}}-E_{i_{1}}\right)  &=&c_{\mu
_{2},Coll}^{{}}\left\langle i_{1}\right| \hat{V}_{1}\left| Coll\right\rangle
+\sum_{i_{2}}c_{\mu _{2},i_{2}}^{{}}\left\langle i_{1}\right| \hat{V}%
_{2}\left| i_{2}\right\rangle =v_{1}c_{\mu _{2},Coll}^{{}}+v_{2}a_{\mu
_{2}}^{2} \\
c_{\mu _{2},i_{2}}^{{}}\left( E_{\mu _{2}}-E_{i_{2}}\right) 
&=&\sum_{i_{1}}c_{\mu _{2},i_{1}}^{{}}\left\langle i_{1}\right| \hat{V}%
_{2}\left| i_{2}\right\rangle ^{*}=v_{2}a_{\mu _{2}}^{1}
\end{eqnarray}
where we have assumed that $\left\langle Coll\right| \hat{V}_{1}\left|
i_{1}\right\rangle =v_{1},$ $\left\langle i_{1}\right| \hat{V}_{2}\left|
i_{2}\right\rangle =v_{2}$ and where we have introduced $a_{\mu
_{2}}^{1}=\sum_{i_{1}}c_{\mu _{2},i_{1}}^{{}},$ $a_{\mu
_{2}}^{2}=\sum_{i_{2}}c_{\mu _{2},i_{2}}^{{}}.$ The first equation gives 
\begin{equation}
c_{\mu _{2},Coll}^{{}}=\frac{v_{1}a_{\mu _{2}}^{1}}{E_{\mu _{2}}-E_{Coll}}
\end{equation}
so that the two other relations reads

\smallskip

\begin{eqnarray}
c_{\mu _{2},i_{1}}^{{}} &=&\frac{1}{E_{\mu _{2}}-E_{i_{1}}}\left( \frac{%
v_{1}^{2}a_{\mu _{2}}^{1}}{E_{\mu _{2}}-E_{Coll}}+v_{2}a_{\mu
_{2}}^{2}\right)  \\
c_{\mu _{2},i_{2}}^{{}} &=&\frac{v_{2}a_{\mu _{2}}^{1}}{E_{\mu
_{2}}-E_{i_{2}}}
\end{eqnarray}
Therefore, the following relations between $a_{1}$ and $a_{2}$ can be
expressed as
\begin{eqnarray}
a_{\mu _{2}}^{2} &=&\sum_{i_{2}}\frac{v_{2}\;a_{\mu _{2}}^{1}}{E_{\mu
_{2}}-E_{i_{2}}}\equiv v_{2}\;a_{\mu _{2}}^{1}\;f_{1}\left( E_{\mu
_{2}}\right)   \label{eq:80} \\
a_{\mu _{2}}^{1} &=&\sum_{i_{1}}\frac{v_{1}^{2}}{E_{\mu _{2}}-E_{i_{1}}}%
\frac{a_{\mu _{2}}^{1}}{E_{\mu _{2}}-E_{Coll}}+\frac{v_{2}a_{\mu _{2}}^{2}}{%
E_{\mu _{2}}-E_{i_{1}}}\equiv \left( \frac{v_{1}^{2}a_{\mu _{2}}^{1}}{E_{\mu
_{2}}-E_{Coll}}+v_{2}\;a_{\mu _{2}}^{2}\right) \;F_{1}\left( E_{\mu
_{2}}\right) 
\end{eqnarray}
where we have introduced the functions $f_{n}$ and $F_{n}$%
\begin{eqnarray}
F_{n}\left( E_{\mu _{2}}\right)  &=&\sum_{i_{1}}\frac{1}{\left( E_{\mu
_{2}}-E_{i_{1}}\right) ^{n}} \\
f_{n}\left( E_{\mu _{2}}\right)  &=&\sum_{i_{2}}\frac{1}{\left( E_{\mu
_{2}}-E_{i_{2}}\right) ^{n}}\;
\end{eqnarray}
Then the eigenenergy $E_{\mu _{2}}$ fulfills the following dispersion
relation 
\begin{equation}
1=\left( \frac{v_{1}^{2}}{E_{\mu _{2}}-E_{Coll}}+v_{2}^{2}\;f_{1}\left(
E_{\mu _{2}}\right) \right) \;F_{1}\left( E_{\mu _{2}}\right) 
\end{equation}
Moreover, the $a$ coefficients can be found using the normalization
condition 
\begin{equation}
1=\frac{v_{1}^{2}a_{\mu _{2}}^{1^{2}}}{\left( E_{\mu _{2}}-E_{Coll}\right)
^{2}}+v_{2}^{2}\;a_{\mu _{2}}^{1^{2}}\;f_{2}\left( E_{\mu _{2}}\right)
+\left( \frac{v_{1}^{2}a_{\mu _{2}}^{1}}{E_{\mu _{2}}-E_{Coll}}+v_{2}a_{\mu
_{2}}^{2}\right) ^{2}\;F_{2}\left( E_{\mu _{2}}\right) 
\end{equation}
which leads to a condition on a, through the use of the relation (\ref{eq:80}%
), in order to remove $a_{\mu _{2}}^{2},$%
\begin{equation}
1=\frac{v_{1}^{2}a_{\mu _{2}}^{1^{2}}}{\left( E_{\mu _{2}}-E_{Coll}\right)
^{2}}+v_{2}^{2}\;a_{\mu _{2}}^{1^{2}}\;f_{2}\left( E_{\mu _{2}}\right)
+\left( \frac{v_{1}^{2}a_{\mu _{2}}^{1}}{E_{\mu _{2}}-E_{Coll}}%
+v_{2}^{2}\;a_{\mu _{2}}^{1}\;f_{1}\left( E_{\mu _{2}}\right) \right)
^{2}\;F_{2}\left( E_{\mu _{2}}\right) 
\end{equation}
allowing to extract the coefficient $a_{1}$%
\begin{equation}
a_{\mu _{2}}^{1^{2}}=\left( \frac{v_{1}^{2}}{\left( E_{\mu
_{2}}-E_{Coll}\right) ^{2}}+v_{2}^{2}\;\;f_{2}\left( E_{\mu _{2}}\right)
+\left( \frac{v_{1}^{2}}{E_{\mu _{2}}-E_{Coll}}+v_{2}^{2}\;f_{1}\left(
E_{\mu _{2}}\right) \right) ^{2}\;F_{2}\left( E_{\mu _{2}}\right) \right)
^{-1}
\end{equation}
Then the overlap matrix, $O_{\mu _{2}}=\left| c_{\mu _{2},Coll}\right| ^{2}$%
, reads

\smallskip 
\begin{equation}
O_{\mu _{2}}=\frac{v_{1}^{2}}{v_{1}^{2}+v_{2}^{2}\;\;f_{2}\left( E_{\mu
_{2}}\right) \left( E_{\mu _{2}}-E_{Coll}\right) ^{2}+\left(
v_{1}^{2}+v_{2}^{2}\;f_{1}\left( E_{\mu _{2}}\right) \left( E_{\mu
_{2}}-E_{Coll}\right) \right) ^{2}\;F_{2}\left( E_{\mu _{2}}\right) }
\end{equation}
Let us now assume as usual that the second level of complexity $\left|
i_{2}\right\rangle $ is characterized by a level spacing $\Delta E_{2},$
then we can write for $f_{1}$%
\begin{equation}
f_{1}\left( E\right) =\frac{\pi }{\Delta E_{2}}\cot \left( \pi \frac{E}{%
\Delta E_{2}}\right) 
\end{equation}
Using the relation (\ref{eq:g_f}), the corresponding relation for $f_{2}$
can be obtained 
\begin{equation}
f_{2}\left( E\right) =\frac{\pi ^{2}}{\Delta E_{2}^{2}}\left( 1+\cot
^{2}\left( \pi \frac{E}{\Delta E_{2}}\right) \right) =\left( \frac{\pi ^{2}}{%
\Delta E_{2}^{2}}+f_{1}^{2}\left( E\right) \right) 
\end{equation}
We get the dispersion relation 
\begin{equation}
f_{1}\left( E_{\mu _{2}}\right) =\frac{F_{1}^{-1}\left( E_{\mu _{2}}\right) 
}{v_{2}^{2}}-\frac{v_{1}^{2}}{\left( E_{\mu _{2}}-E_{Coll}\right) v_{2}^{2}}
\end{equation}
then we get for the coupling coefficient $c_{i_{1}}^{\mu _{2}}$ 
\begin{equation}
O_{\mu _{2}}=\left| c_{\mu _{2},Coll}^{{}}\right| ^{2}=\frac{%
v_{2}^{2}v_{1}^{2}}{v_{2}^{2}v_{1}^{2}+\frac{\Gamma _{2}^{2}}{4}\left(
E_{\mu _{2}}-E_{Coll}\right) ^{2}+\left( \frac{\left( E_{\mu
_{2}}-E_{Coll}\right) }{F_{1}\left( E_{\mu _{2}}\right) }-v_{1}^{2}\right)
^{2}+\;\frac{\left( E_{\mu _{2}}-E_{Coll}\right) ^{2}v_{2}^{2}F_{2}\left(
E_{\mu _{2}}\right) }{F_{1}^{2}\left( E_{\mu _{2}}\right) }}
\end{equation}
where we introduce $\Gamma _{1}=2\pi v_{2}^{2}/\Delta E_{2}$. If we assume
that the interaction $v_{2}^{2}\rightarrow 0$ we can approximate the overlap
by 
\begin{equation}
O_{\mu _{2}}=\left| c_{Coll}^{\mu _{2}}\right| ^{2}=\frac{v_{2}^{2}v_{1}^{2}%
}{\frac{\Gamma _{1}^{2}}{4}\left( E_{\mu _{2}}-E_{Coll}\right) ^{2}+\left( 
\frac{E_{\mu _{2}}-E_{Coll}}{F_{1}\left( E_{\mu _{2}}\right) }%
-v_{1}^{2}\right) ^{2}}
\end{equation}
Let us now consider the case of a regular ensemble of states $i_{1}$ with a
spacing $\Delta E_{1}$, we can write for $F_{1}$%
\begin{equation}
F_{1}\left( E_{\mu _{2}}\right) =\frac{\pi }{\Delta E_{1}}\cot \left( \pi 
\frac{E_{\mu _{2}}}{\Delta E_{1}}\right) 
\end{equation}
and the overlap reads 
\begin{equation}
O_{\mu _{2}}=\left| c_{Coll}^{\mu _{2}}\right| ^{2}=\frac{v_{2}^{2}v_{1}^{2}%
}{\frac{\Gamma _{1}^{2}}{4}\left( E_{\mu _{2}}-E_{Coll}\right) ^{2}+\frac{^{%
\Delta E_{1}^{2}}}{\pi ^{2}}\left( \left( E_{\mu _{2}}-E_{Coll}\right) \tan
\left( \pi \frac{E_{\mu _{2}}}{\Delta E_{1}}\right) -\frac{\Gamma _{Coll}}{2}%
\right) ^{2}}
\end{equation}
where we have introduced $\Gamma _{Coll}=2\pi v_{1}^{2}/\Delta E_{1}$.

\end{document}